# CONTROL BASED ENHANCED REGENERATIVE MODES FOR HYDRAULIC MULTI-ACTUATOR SYSTEMS

**BIDEAUX Eric**
Laboratoire Ampère, Université de Lyon, INSA Lyon
Ecole Centrale de Lyon, Université Claude Bernard
Lyon 1, CNRS, Campus de LA DOUA, 69 621
Villeurbanne, France
Email: eric.bideaux@insa-lyon.fr

**TARDY Grégory**
Laboratoire Ampère, Université de Lyon, INSA Lyon
Ecole Centrale de Lyon, Université Claude Bernard
Lyon 1, CNRS, Campus de LA DOUA, 69 621
Villeurbanne, France
Volvo Compact Equipment, 01 300 Belley, France

**SMAOUI Mohamed**
Laboratoire Ampère, Université de Lyon, INSA Lyon
Ecole Centrale de Lyon, Université Claude Bernard
Lyon 1, CNRS, Campus de LA DOUA, 69 621
Villeurbanne, France

**LANOIX Aurore**
Cerebrum Ingénierie, 42 420 Lorette, France

## ABSTRACT

This paper focuses on a control-based approach for enhancing the regenerative capabilities of hydraulic multi-actuator systems using individual metering valves. Thanks to this architecture, pressure and displacement of each actuator can be controlled nearly independently. By determining online, the right pressure to be driven, it enables the optimization of regenerative control strategies for resistive or driving forces. Globally, this control strategy behaves such as a load sensing approach but each metering valve is piloted in order to activate regenerative mode when it is allowable.
The main contribution relies on optimizing the pressure to be controlled in each actuator and the main pump in order to maximize the regenerative capacity of a hydraulic machine while following a displacement. The effectiveness of the proposed approach is proved in simulation. Only a single pump line regeneration is explored here but extensions to multi-pump or direct regeneration are also possible.

## KEYWORDS

## I INTRODUCTION

Electrification is a major shift for reducing emission of mobile machines and it benefits of the previous innovation in the car and truck industry. However, autonomy and Total Cost of Ownership (TCO) of these electrified machines remain real challenges since conventional hydraulic circuit topologies and machine control approaches are not really designed in order to maximize the global efficiency. Improving the hydraulic efficiency about ten percent has exactly the same impact on autonomy or Total Cost of Ownership (TCO). The current context of global warming and increasingly strict anti-pollution regulations is driving the heavy-duty mobile machine industry towards carbon-free powertrain solutions such as electric or fuel-cell ones to power their new machines. The low energy density of electric batteries and dihydrogen tanks compared to diesel fuel brings new challenges in terms of machine design: carrying energy is bulky, heavy and expensive in the case of electric batteries. Thus, to ensure the carbon-free powertrain transition on heavy-duty mobile machines it is necessary to enhance the transmission energy efficiency.

The literature on heavy-duty mobile machine based on electrohydraulic technologies is quite extensive (Padovani et al., 2024) with different concepts such as independent-metering (Eriksson et al., 2011 and Vukovic et al., 2014), digital hydraulics (Linjama 2011, Scheidl 2012) and pump speed or displacement-controlled systems (Fassbender, 2021). All these concepts differ in terms of components and control strategies, they can be loosely categorized from multiple valves, complex control and low-price tag systems to multiple pumps/electric motors, valveless and expensive systems. By looking deeper through the hydraulic systems literature, the different concepts definitions appear quite porous, many systems are found to be hybridisation of different concepts.

Qu et al. (2023) proposed electrohydraulic units with a valve-controlled bypass allowing for more efficient control on low speeds actuator operations. A more general definition of electrohydraulic systems that covers systems ranging from individual metering to pump controlled concepts can be intuited for systems with one or more pumps and multiple valves (Weber, 2016). In Vukovic et al. (Vukovic, 2013) a general hydraulic system definition with such a viewpoint is introduced. Three main control approaches were defined,

Primary, Conductive and Secondary control related to the different actuated components for control: the hydraulic power sources, the distribution components or the hydraulic actuators. These control approaches can be mixed benefiting from all available degrees of freedom on an optimized electrohydraulic system and operating points to reach the highest energy efficiency. While Larsson et al. (Larsson, 2021) formulated the optimal sizing and control of a digital displacement pump as a Dynamic Programming (DP) approach, a more general approach is proposed by Tardy et al. (Tardy, 2024) using a Mixed Integer Non-Linear Programming (MINLP).

The literature on electrohydraulic systems control is often focused on simulations and/or tests of online rule-based control strategies (Sidhom, 2009), some of these strategies use specific metering modes called regeneration or recuperation with the intent of benefiting from high efficiency flow paths. Electrohydraulic systems often provide many degrees of freedom making the optimal metering computation a complex task. Multivariable nonlinear control is the core of this approach (Schaep, 2015). It has already been extensively studied in the fluid power context, but rarely applied in practice. With electrification and individual metering valves, it makes its implementation judicious and affordable if the number of sensors is reasonable.

At the control synthesis stage, whereas it is straightforward to define the user desired position/velocity as one of the degrees of freedom (DoF), the second DoF can be related to various objectives (Xu 2013,). Several papers explore this opportunity for reducing the energy consumption, for example by setting a low pressure (Jung 2022, Ketonen 2017). However, even this could be relevant in the case of a single actuator, prioritizing regenerative modes (Rydberg, 2005) can be almost as favourable, especially in the context of multiple actuators.

The main contribution of the paper relies on defining the pressure to be controlled in each actuator in order to maximize the regenerative capacity of a hydraulic machine. Of course, regeneration is only admissible in some cases that depend on the displacement direction and the force applied on the cylinder rod. Therefore, a sliding mode observer (SMO) is developed in order to determine with a minimum number of sensors the disturbance force on each actuator and feed the regenerative mode selection algorithm as well as the variable pressure trajectory control law. In this paper, only a single pump line regeneration is explored but extension to multi-pump or direct regeneration is possible.

Following this introduction, a multivariable nonlinear control will be described in the next section based on a generic hydraulic actuation model. In section III, the algorithm proposed for prioritizing the regenerative modes when admissible and favourable, is detailed before introducing the force disturbance SMO in section IV. Simulation results will underline the achieved performances in terms of observer performance, regenerative mode activation and efficiency, and also energy saving in section V. Finally, conclusion will be drawn concerning real-world hydraulic applications and contribution of the proposed approach to the advancement of sustainable and energy-efficient fluid power technologies.

# II SYSTEM MODELLING AND CONTROL

Figure 1 illustrates a common hydraulic architecture, which consists of a set of $n$ cylinders supplied by a single variable displacement pump that drives a variable supply pressure line $p_S$ with a flow rate $Q_S$. Each actuator $i$ is piloted by a set of independent meter-in and meter-out valves. The valve dynamic is not modelled and its opening section $s_{vxi}$ is taken as a function of the command $u_{ji}$.

Each actuator $i$ is modelled using usual assumptions. Dry friction phenomena are included in the disturbance force term. Pipes and other non-functional pressure drop in the circuit as well as leakages are not taken into account in the model for control purposes. Without loss of generality the control inputs are the flow rates $Q_{ai}$ and $Q_{bi}$.

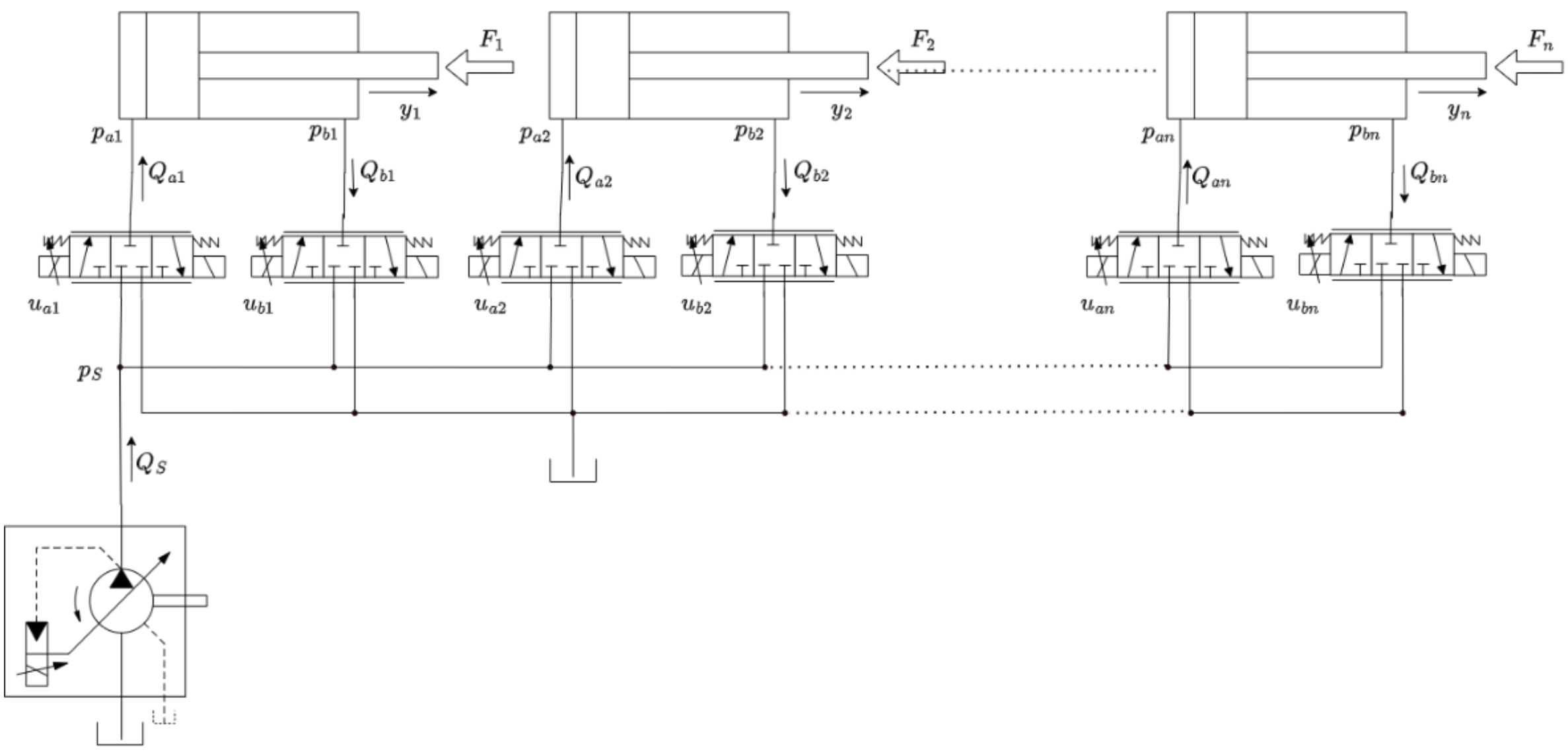

*Figure 1:* Multi-actuator hydraulic architecture

$$\begin{cases} \frac{d}{dt}p_{ai} = \frac{B}{V_{ai}(y_i)}(Q_{ai} - S_{ai}v_i) \\ \frac{d}{dt}p_{bi} = \frac{B}{V_{bi}(y_i)}(-Q_{bi} + S_{bi}v_i) \\ \frac{d}{dt}v_i = \frac{1}{M}(S_{ai}p_{ai} - S_{bi}p_{bi} - b_i v_i - F_i) \\ \frac{d}{dt}y_i = v_i \end{cases} \quad (1)$$

$$Q_{xi} = c_s s_{vxi}(u_{xi})\sqrt{\frac{2}{\rho}\Delta p_{xi}}\,\mathrm{sign}(\Delta p_{xi}) \quad (2)$$

with $\Delta p_{xi} = \begin{cases} p_S - p_{xi}, & u_{xi} \geq 0 \\ p_{xi}, & u_{xi} < 0 \end{cases}, x \in \{a, b\}$

By differentiating the velocity, the model of actuator $i$ can be written as:

$$\begin{cases} \frac{d}{dt}y_i = v_i \\ \frac{d}{dt}v_i = a_i \\ \frac{d}{dt}a_i = -\frac{B}{M}\left[\frac{S_{ai}^2}{V_{ai}(y_i)} + \frac{S_{bi}^2}{V_{bi}(y_i)}\right]v_i + \cdots \\ +\frac{B}{M}\left[\frac{S_{ai}}{V_{ai}(y_i)}Q_{ai} + \frac{S_{bi}}{V_{bi}(y_i)}Q_{bi}\right] - \frac{b}{M}a_i - \frac{1}{M}\frac{d}{dt}F_i \\ \frac{d}{dt}p_{bi} = \frac{B}{V_{bi}(y_i)}(-Q_{bi} + S_{bi}v_i) \end{cases} \quad (2)$$

Thanks to the 2 degrees of freedom of the command, if we consider the position $y_i$ and the pressure $p_{bi}$ as the 2 outputs, it can easily be shown that the characteristic index of the model is 4, that is the dimension of the system. This means that all the system dynamics can be controlled. Moreover, rod displacement and one of the cylinder chamber pressures can be piloted independently. This property is essential to be able to develop regenerative mode control. Let us note that if a single 4/3 valve is used, the pressure dynamics would have been uncontrollable.
The model may then be expressed as input-affine nonlinear model and according to the characteristic index of the system, a full nonlinear state feedback linearization can be proposed. This leads to the following formulation:

$$\begin{cases} \dot{x}_i = f_i(x) + g_i(x)u_i + g_2(x)w_i \\ \dot{z}_i = f_{zi}(x) + g_{zi}(x)u_i \end{cases} \quad (3)$$

with $x_i = [y_i, v_i, a_i]^t$, $z_i = [p_{bi}]^t$, $u_i = [Q_{ai}, Q_{bi}]^t$, $w_i = \frac{1}{M}\dot{F}_i$

Let us define the new commands $u_{xi}$ and $u_{zi}$ as:

$$Q_{bi} = -\frac{V_{bi}(x_{1i})}{B}u_{zi} + S_{bi}x_{2i} \quad (4a)$$

$$Q_{ai} = \frac{V_{ai}(x_{1i})}{S_{ai}}\left[\frac{M}{B}u_{xi} + \frac{S_{bi}}{B}u_{zi} + \frac{S_{ai}^2}{V_{ai}(x_{1i})}x_{2i} + \frac{b}{B}x_{3i}\right] \quad (4b)$$

Then, the model can be fully linearized, all the nonlinearities being included in the new commands. It can be written as:

$$\begin{cases} \dot{x}_i = \begin{bmatrix} 0 & 1 & 0 \\ 0 & 0 & 1 \\ 0 & 0 & 0 \end{bmatrix} x_i + \begin{bmatrix} 0 \\ 0 \\ 1 \end{bmatrix} u_{xi} + \begin{bmatrix} 0 \\ 0 \\ 1 \end{bmatrix} w_i \\ \dot{z}_i = u_{zi} \end{cases} \quad (5)$$

In these 2 independent models, the 3 first equation corresponds to the position tracking model, while the last equation is the pressure control model. Accordingly, the control of actuator $i$ can be split in two separated parts: the position and the pressure controllers.
The position model is a cascade of 3 integrators and therefore the controller synthesis can be achieved using different techniques, which have already extensively studied in the literature (Xu 2013, Linjama 2016) such as sliding mode, backstepping, and so on. To make it simple, a pole placement is used with an additional position error integral term, which enable the modelling errors and the disturbances to be rejected. This controller is defined as

$$u_{xi} = k_{x0}\int_t \varepsilon_{x1}\,d\tau + k_{x1}\varepsilon_{x1} + k_{x2}\varepsilon_{x2} + k_{x3}\varepsilon_{x3} \quad (6)$$

where $\varepsilon_{xi} = \left(x_i - x_i^{\#}\right)$

For the pressure controller, a P- or PI-controller can be used

$$u_{zi} = k_{z1}\varepsilon_z + k_{z2}\int_t \varepsilon_z\,d\tau \quad (7)$$

It is important to note that the objective of the control synthesis presented in this section is to demonstrate the system's controllability and to highlight the separation between position and pressure control objectives. In practice, only the pressure control loop is of interest since the position control is achieved by the driver, who is generally closing the loop. However, it can be applied in the case of automatic tasks with a predefined movement. Nevertheless, in this paper, for simulation purpose the position tracking controller will enable the desired trajectory to be nicely followed.

# III OPTIMAL OPERATING PRESSURES

Energy recovery can be achieved when part of the mechanical power can be recovered from an actuator and be converted into hydraulic power that can be directly used or stored in order to reduce the required pump energy over a cycle. Actually, braking energy is rarely recovered in hydraulic systems while most of the components are reversible like electric drives. This participates to the low efficiency of hydraulic circuits.
In multi-actuator systems, regenerative modes can be activated in different conditions according to the topology of the circuit and the operating conditions (Tardy 2024). Of course, the circuit architecture can be designed to optimize flow paths to be piloted between actuators, intermediary pressure level lines can be defined with or without accumulators (Schaep, 2015), several pumps or pump/motors (Vaezi 2020) can be used. The sizing of the actuators is also a degree of freedom that can also be explored.

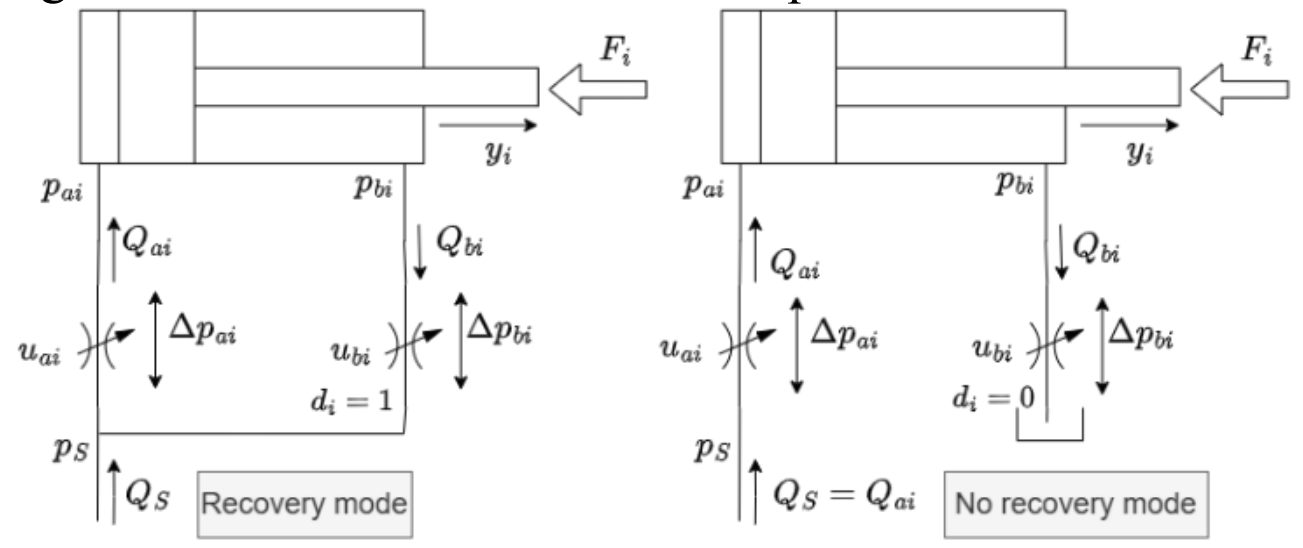

*Figure 2:* Recovery mode for a single actuator

In this paper, we consider the simplest case, in which all the actuators are supplied by a single pressure-controlled pump and exhausts are directed to the tank. Therefore, energy recovery is only realizable through the supply line at the pump pressure $p_p$. Thanks to the metering valves on each actuator, one of the chamber pressures can be controlled independently from the piston displacement and can be set to activate energy recovery if some conditions are reached. These conditions can be calculated analytically but for complex systems with many flow paths, an optimization algorithm can efficiently address this problem and determine the optimal operating pressure for each cylinder and for the pump.

Let us consider the example of figure 2 where we assume that $v_i > 0$. In steady state conditions, i.e. $v_i = cte$ and $F_i = cte$, the regenerative mode can be activated ($d_{bi} = 1$) if there exist pressures that verify

$$\begin{cases} S_{ai}p_{ai} - S_{bi}p_{bi} - F_i = 0 \\ p_{ai} + \Delta p_{ai} = p_S \\ p_{bi} = p_S + \Delta p_{bi} \end{cases} \tag{8}$$

In non-recovery mode ($d_{bi} = 0$), these constraints become

$$\begin{cases} S_{ai}p_{ai} - S_{bi}p_{bi} - F_i = 0 \\ p_{ai} + \Delta p_{ai} = p_S \\ p_{bi} = \Delta p_{bi} \end{cases} \tag{9}$$

Using a binary variable, a conditional constraint can be introduced to cover both cases as follows

$$p_{bi} = d_{bi}p_S + \Delta p_{bi} \tag{10}$$

Since the flow rate in each chamber is imposed by the piston velocity, the pressure drops across each valve have to respect the minimum pressure at full opening, it leads to the additional constraints

$$\begin{cases} \Delta p_{ai} \geq \frac{\rho}{2C_dS_{vai}} Q_{ai}^2 \\ \Delta p_{bi} \geq \frac{\rho}{2C_dS_{vbi}} Q_{bi}^2 \end{cases} \tag{11}$$

Activating the regenerative mode leads to increase the pressure opposed to the displacement and therefore the supply pressure. Then, in order to verify that the regenerative mode is energetically relevant, the optimization problem will aim at minimizing the pump power define by the cost function

$$J(p_S, d_{bi}, v_i) = p_S(S_{ai} - d_{bi}S_{bi})v_i \tag{12}$$

This approach can be extended to all the actuators and velocity sign and define a global optimization problem that may be formulated as a Mixed Integer Linear Problem (MINLP):

$$\left| \begin{array}{l} J^* = arg \min_{\chi} J(p_S, d_{bi}, d_{ai}, p_{ai}, p_{bi}) \\ \quad st. \;\; A.\chi \leq 0 \\ \quad\quad\;\; B(d_{bi}, d_{ai}).\chi = 0 \end{array} \right. \quad \text{with} \;\; \chi = (p_S, p_{ai}, p_{bi}) \in [0, p^{max}]^3 \tag{13}$$

For illustrating the results of this algorithm, let us consider an academic example with 2 cylinders. Figure 3 shows the velocity of the cylinders and the force applied on the rods. Here, sine wave functions are used in order to cover most of the cases. As shown in figure 5, the recovery mode is activated mostly for negative power, that is for opposite sign of effort and velocity. Figure 4 shows the optimal operating pressures in chamber A and B for both cylinders.

Until t=3 s, cylinder 2 is in recovery mode and flow from chamber B is recovered at chamber A of the same cylinder. This is energetically efficient since the minimum value of the supply pressure $p_S$ is set by cylinder 1 and the flow recovery reduces the flow to be supplied by the pump with about 5 Mpa pressure drop. Then, at about t=9 s, cylinder 2 switches again in recovery mode and until t=11 s this energy is used either in chamber B of cylinder 1 or in chamber A of cylinder 2. The supply pressure $p_S$ follows this value but provide only a little flow. Then, between t=11 s and t=13 s, the pressures at port B for both cylinders are set in order to brake the displacement due to the high driving force but this flow is almost lost since the supply pressure $p_S$ is set to the minimum value. After t=13 s and until t=17 s, only cylinder 1 stays in recovery mode and recovered energy is used to pressurized the chambers of cylinder 2.

Due to the linear form of the constraints, the optimization problem can be solved very efficiently and implemented online. It requires the force applied on each actuator and the velocity sign for each piston, and provides the pressure reference to be tracked in each actuator and at pump. While the velocity can easily be obtained from the driver joystick, a direct measurement of the applied force is non-realistic. Therefore, in the next section, a sliding mode observer of the applied force is proposed in order to avoid a direct measurement.

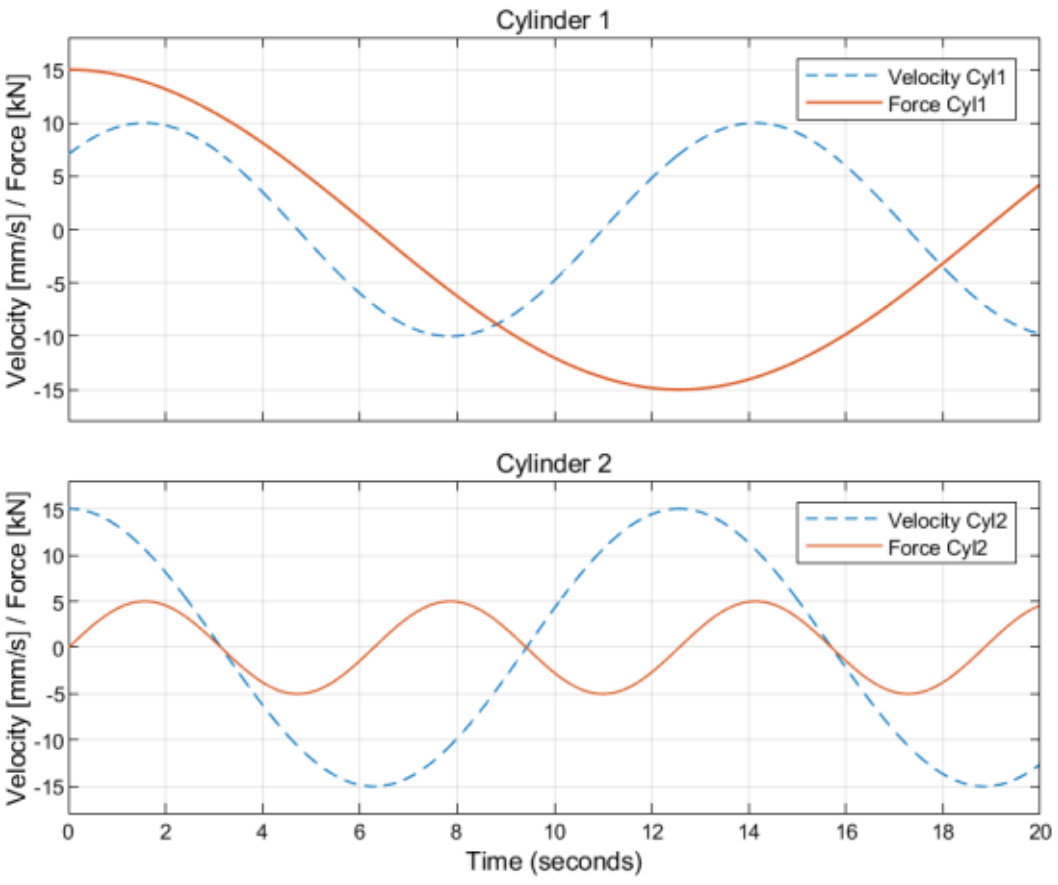

*Figure 3:* Piston velocity and applied external force for each cylinder

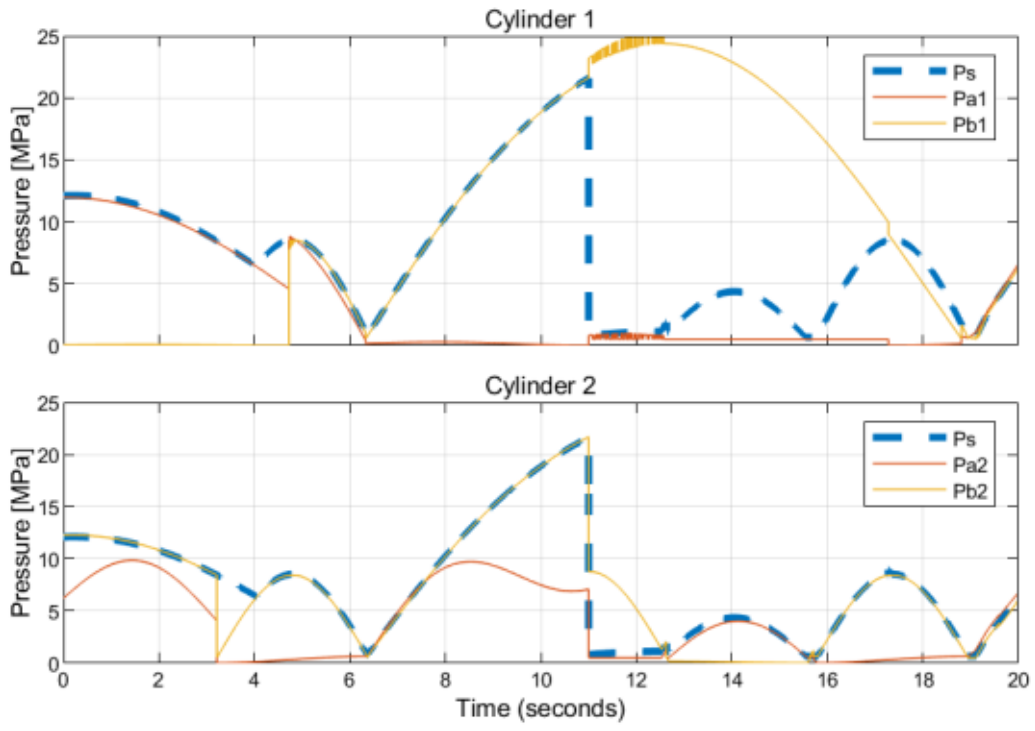

*Figure 4:* Operating pressure optimization results

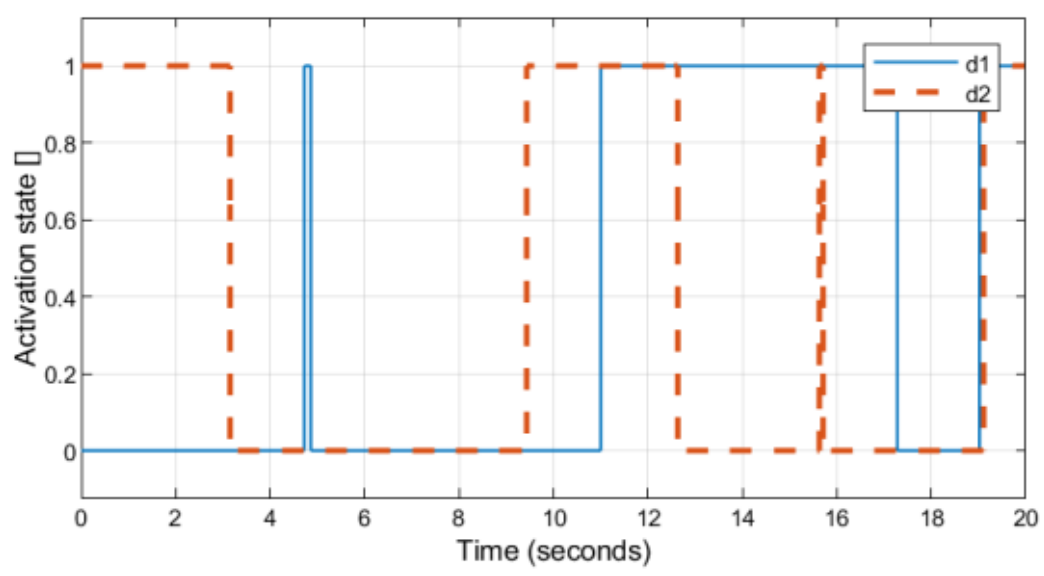

*Figure 5:* Regenerative mode activation states

# IV DISTURBANCE FORCE OBSERVER

Many schemes for the estimation of states variables have been proposed in recent years. Some of these methods are based on nonlinear observer theory such as high gain, sliding mode and backstepping observers. However, nonlinear state observers are difficult to implement when poor knowledge on the system dynamics is available. Therefore, research on modelling and identification is necessary to improve the performance of observers. Another attractive method for the estimation of states variables, especially for mechanical systems, is the numerical differentiation. Indeed, Differentiators are very useful tools to determine and estimate signals. For example, for mechanical systems, the velocity and acceleration can be computed from the position measurements using differentiators. However, the design of an ideal differentiator is a difficult and a challenging task. Ibrir (2004) discusses the properties and the limitations of two different structures of linear differentiation systems. A predictive algorithm which is applied to angular acceleration measurements is presented by Valiviita (1998).

In order to estimate the force applied on the cylinders, we propose here a robust differentiator via sliding mode technique. Indeed, a robust exact differentiation via sliding mode technique is proposed in Levant (1998). The differentiator considered features simple form and easy design.

Without loss of generality, let input signal $f(t)$ be a measurable function and let it consist of a base signal having a derivative with Lipschitz's constant $C > 0$. In order to differentiate the input signal, consider the auxiliary equation

$$\dot{x} = u \tag{14}$$

Consider now the following sliding surface which represent the difference between $x$ and $f(t)$

$$s = x - f(t) \tag{15}$$

By differentiating s, it leads to the following relationship

$$\dot{s} = u - \dot{f}(t) \tag{16}$$

The super twisting algorithm introduced by Levant (1998) defines the control law $u$ as

$$u = u_1 - \lambda |s|^{\frac{1}{2}} sgn(s) \tag{17}$$

with

$$\dot{u_1} = -w\, sgn(s) \tag{18}$$

where $w, \lambda > 0$

Here $u$ is the output of the differentiator. Indeed, the super twisting algorithm converges in finite time, so the following relationship can be obtained in finite time:

$$\dot{x} - \dot{f(t)} = u - \dot{f(t)} = 0 \tag{19}$$

or

$$u = \dot{f(t)} \tag{20}$$

The corresponding sufficient conditions for finite time convergence are:

$$w > C \tag{21}$$

$$\lambda^2 \geq 4C \, \frac{w+C}{w-C} \tag{22}$$

The separation principle is fulfilled for the proposed differentiator (Figure 6). For the design purpose consider the dynamics of the mechanical part of the cylinder expressed as (1), where the position $y_i$ and the pressures $p_{ai}$ and $p_{bi}$ are measured. In order to reconstruct the bounded external load force $F_i$,the robust differentiator via sliding mode technique is used to recover the velocity and the acceleration signals from the position (Sidhom 2014). Consequently, the external force is recovered by:

$$\widehat{F}_i = S_{ai} p_{ai} - S_{bi} p_{bi} - b\widehat{v}_i - M\widehat{a}_i \tag{23}$$

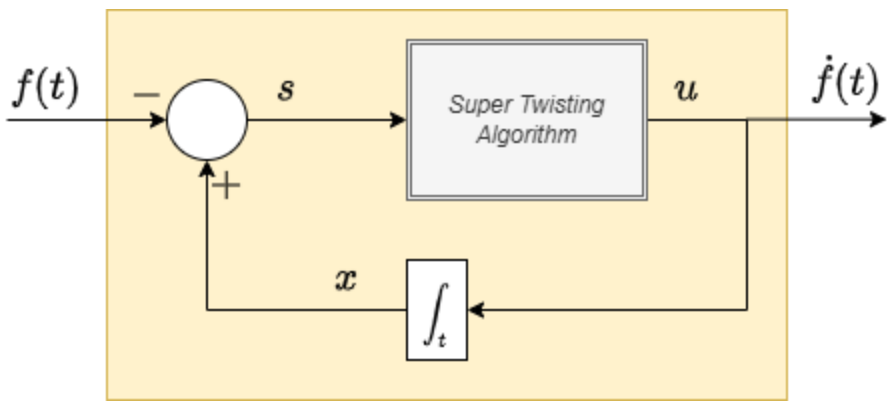

*Figure 6:* Structure of the differentiator

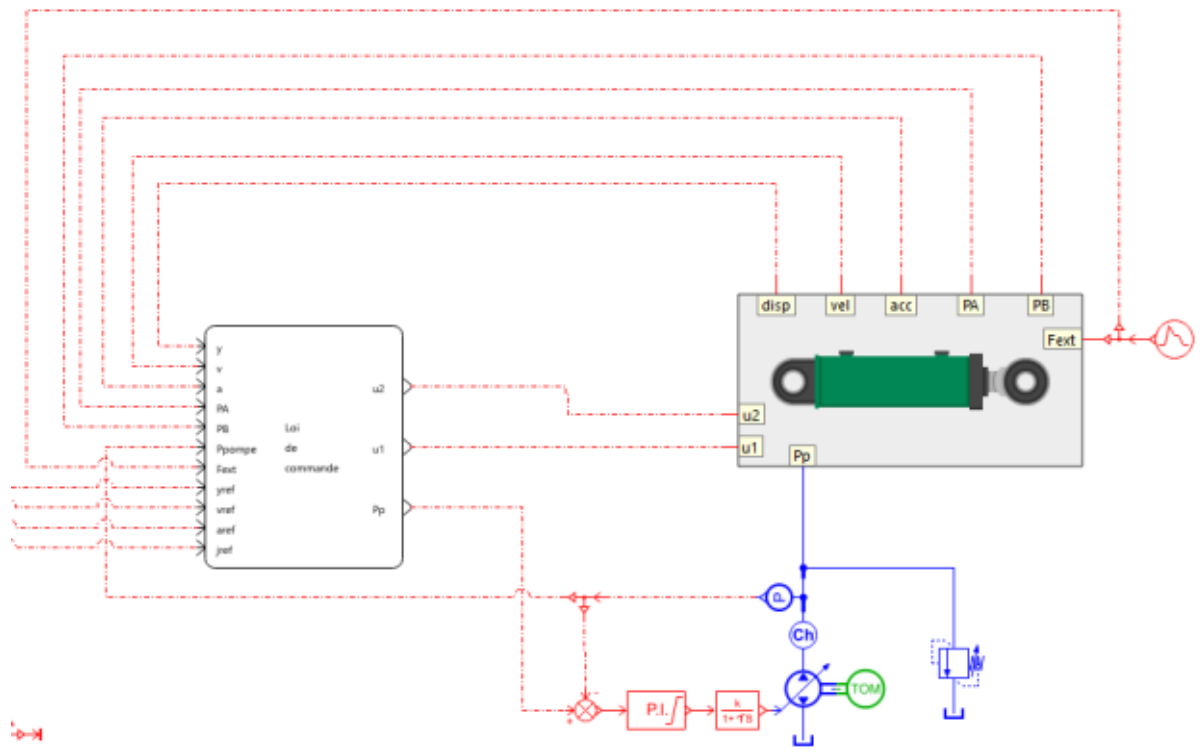

*Figure 7:* Co-simulation model

## V SIMULATION

The performance of the whole algorithm is illustrated by a co-simulation between Matlab/Simulink and LMS Simcenter AMESim (Figure 7). For sake of simplicity, the system studied here consists in a single actuator following a sine wave displacement trajectory under a force disturbance (Figure 8).

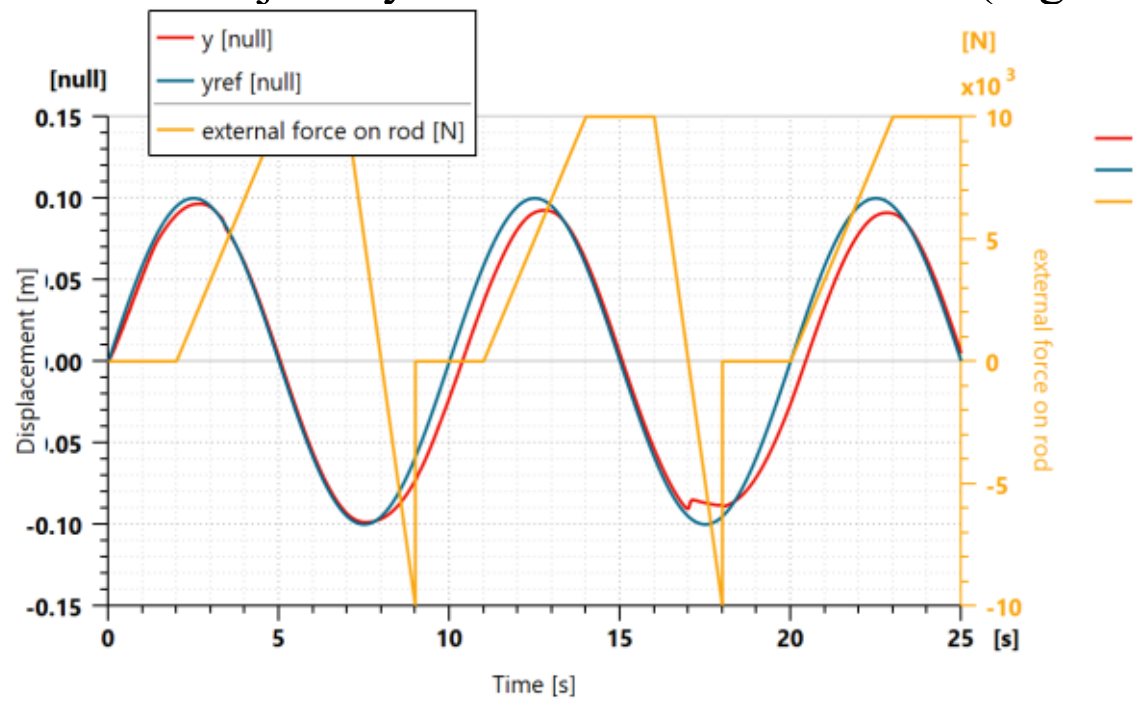

*Figure 8:* Reference displacement trajectory and disturbance force applied on rod

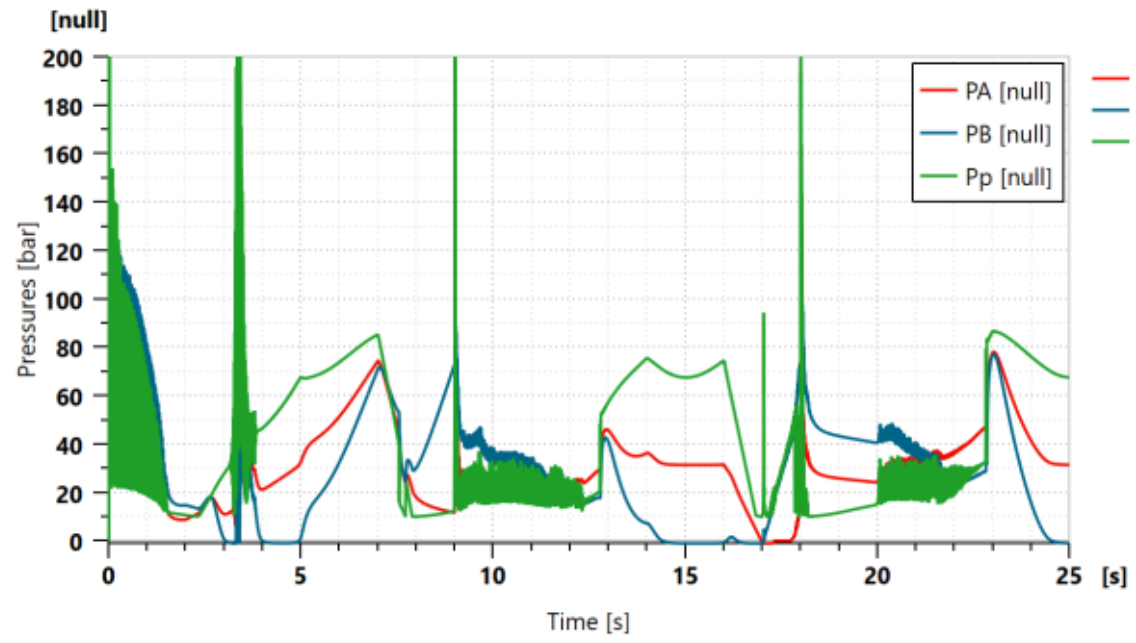

*Figure 9:* Pressures at port A and B of the cylinder and pump pressure

The model of the system developed in LMS Simcenter AMESim aims at representing a realistic case. It includes many details such as dynamics of proportional valves (80 Hz) and pump displacement (17 Hz), dry friction and leakages in cylinder, and some uncertainties compared to the control model such as valve flow characteristics, cylinder dead volumes, viscous friction coefficient, etc.
The control block includes the different parts of the control law. Among them, the optimization algorithm provides the pump and chamber B pressure setting points that enables optimal recovery behaviour. It uses the SMO of the force required on each actuator and the velocity measurement. Finally, the nonlinear control for trajectory tracking and the chamber B pressure control that uses the imposed displacement trajectory and the reference pressure (pump and chamber B) for calculating the commands of the 2 proportional valves.
As shown on Figure 8, the trajectory is tracked nicely even under the force disturbance and the pump pressure variations. The linear part of the controller (6) may certainly be improved in order to reach a better precision; however, as stated previously, this is achieved by the driver in practice and automatic tasks are not the purpose here. Figure 9 shows the pressures at port A and B of the cylinder. The pressure at port B follows nicely the optimal pressure defines by the optimization algorithm and enables energy recovery when adequate conditions are reached. The activation of recovery modes can be identified by zones where the pump pressure is lower than the pressure at port B, that is around 10 s and 20 s. Of course, this case of study with a single actuator is less favourable than a multi-actuator system.

# V CONCLUSION

In this paper, we developed a novel approach that aims at improving the energy efficiency of a muti-actuator system using individual metering valves for each chamber of the cylinders. Based on a nonlinear control approach that enables decoupling the model of each actuator, the displacement trajectory and the pressure in one chamber can be tracked in order to introduce recovery modes within the system. An optimization algorithm has been proposed and enables the optimal operating pressures for the cylinders and the pump to be determined and used as references to be tracked by the control. It requires the force required by each actuator to be estimated and a SMO is developed for this purpose.
Simulation results show that both the optimization, the proposed control and observer allow recovery modes to be activated when conditions are adequate. Future work will aim at improving the control and the optimization algorithms in order to reduce fast dynamics and oscillatory behaviour that appears when recovery modes are activated. The drivability of the system is also an essential aspect to be explored since such variable pump pressure can make the control more difficult from the driver point of view. More realistic cases have also to be studied in order to evaluate the energy gain and the effectiveness of this approach.

# NOTATIONS

Variables and parameters

| | |
|---|---|
| $B$ | Bulk modulus (Pa) |
| $p$ | Pressure (Pa) |
| $V$ | Actuator chamber volume (m$^3$) |
| $S$ | Actuator section (m$^2$) |
| $Q$ | Flow rate (m$^3$/s) |
| $c_d$ | Discharge coefficient |
| $\Delta p_{ai}$ | Flow rate (m$^3$/s) |
| ρ | Fluid density (kg/m$^3$) |
| $s_v$ | Valve nominal section (m$^2$) |
| $b$ | Viscous friction coefficient (N.s/m) |
| $F$ | Force (N) |
| $y$ | Displacement (m) |
| $v$ | Velocity (m/s) |
| $a$ | Acceleration (m/s$^2$) |
| $j$ | Jerk (m/s$^3$) |
| $x$ | State vector |
| $u$ | Input vector |
| $w$ | Disturbance vector |
| $k_{xi}, k_{zi}$ | Linear corrector parameters |
| $\varepsilon_{xi}, \varepsilon_{zi}$ | Errors |
| $\lambda, C, w$ | Sliding mode observer constants |
| $t$ | Time (s) |

Exponents and indices

| | |
|---|---|
| $a_i$ $(or\ b_i)$ | index for actuator *i* port A (or B) |
| $s$ | index for pump (supply) |
| # | index for reference value |